\documentclass[epj]{svjour}
\usepackage{graphics}
\usepackage{amsmath}
\usepackage{graphicx}
\usepackage{amssymb}
\begin{document}

\title{Entropic stochastic resonance:
  the constructive role of the unevenness}
\author{P.S. Burada\inst{1}, G. Schmid\inst{1}, D. Reguera\inst{2},
  J.M. Rubi\inst{2} \and P. H\"anggi\inst{1}}
\institute{
  Institut f\"ur Physik, Universit\"at Augsburg, Universit\"atsstr. 1,
  D-86135 Augsburg, Germany
  \and
  Departament de F\'isica Fonamental, Facultat de F\'isica,
  Universidad de Barcelona, Mart\'i i Franqu\'es 1, 08028 Barcelona, Spain}
\date{Received: date / Revised version: date}
%
\abstract{We demonstrate  the existence of stochastic resonance (SR)
in confined systems arising from entropy variations associated to
the presence of irregular boundaries.  When the motion of a Brownian
particle is constrained to a region with uneven boundaries, the
presence of a periodic input may give rise to a peak in the spectral
amplification factor and therefore to the appearance of the SR
phenomenon.  We have proved that the amplification factor depends on
the shape of the region through which the particle moves and that by
adjusting its characteristic geometric parameters one may optimize
the response of the system. The situation in which the appearance of
such entropic stochastic resonance (ESR) occurs is common for
small-scale systems in which confinement and noise play an prominent
role. The novel mechanism found could thus constitute an important
tool for the characterization of these systems and can put to use
for controlling  their basic properties.
}
\authorrunning{Burada {\itshape et al.}}
\titlerunning{Entropic Stochastic Resonance}

\maketitle

\section{Introduction}

A Brownian particle moving in a  threshold-like potential landscape
and subjected to the influence of a periodic forcing may exhibit a
coherent response giving rise to an amplification of the input at a
certain optimal value of the noise level. This resonant phenomenon,
observed in general in the wide class of periodically modulated
noisy systems, was termed stochastic resonance and constituted a
paradigm shift in the way we think about noise effects in systems
away from equilibrium \cite{gammaitoni}. In this new paradigm, the
presence of noise does not always constitute a nuisance; on the
contrary, it may play a constructive role
\cite{gammaitoni,bulsaraJS,PT_SR,chemphyschem,vilar_mono,Lutz99,Schmid01,BuchSR,Yasuda08,scholarpediaSR,GoychukSRa,GoychukSRb}.

Up to now, the phenomenon of SR has been observed mainly in systems
dominated by the presence of a purely energetic potential or
possessing some dynamical threshold \cite{gammaitoni}. However, when
scaling down the size of a system, the free energy rather than the
internal energy becomes the most appropriate potential, and there
are cases in which changes in the free energy are mainly due to
entropy variations
\cite{GoychukSRa,hille,zeolites,liu,berzhkovski,Reguera_PRL,Burada_PRL}.
This is what occurs in constrained systems. In the case of a
Brownian particle moving in a confined medium, entropy variations
contribute to changes in the free energy and may under some
circumstances become its leading contribution
\cite{GoychukSRb,hille,Reguera_PRL,Burada_PRL}. We will show in this
work that the unevenness may also give rise to a stochastic
resonance effect and that this effect can be controlled upon
variation of the geometrical parameters, characterizing the shape of
the cavity in which the Brownian particle dwells.

Usually, the analysis of SR effects have been performed by means of
pertinent Langevin or corresponding Fokker-Planck models
\cite{hanggithomas,Risken}. In confined systems the presence of
boundaries exerts a strong influence in the dynamics and one has to
solve the corresponding boundary value problem. This task cannot
always be easily achieved. The fact that in many instances
boundaries are very intricate enormously complicates the
mathematical treatment of the problem to the extent of becoming a
Herculean task when the boundaries are extremely irregular
\cite{Burada_PRE}. This feature demands the implementation of
different approaches entailing a simplification of the analysis
\cite{Jacobs,Zwanzig,Reguera_PRE}. Among them, the Fick-Jacobs
equation, based on a coarsening of the description in terms of a
single, relevant coordinate degree of freedom, accurately performs
this task \cite{Reguera_PRL,Burada_PRL,Burada_PRE,Jacobs,Zwanzig}.
This methodology will guide us in this article to analyze the
appearance of the SR effect in presence of unevenness.

The article is organized in the following way. In Section 2,
we introduce a model for Brownian motion in a confined medium.
In Section 3, we present a reduction method which simplifies the
complex nature of the 3D/2D dynamics giving rise to an
effective one-dimensional kinetic description.
Section 4 is devoted to evaluate the transition rate from the reduced kinetic description,
and the introduction of spectral amplification
within the two-state approximation.
In Section 5, we present the results showing the ESR phenomenon.
The impact of the geometrical shape and confinement on the
spectral amplification is discussed in Section 6.
Finally, we summarize our main conclusions in Section 7.

\section{Confined Brownian motion}
\label{sec:cbm}

\begin{figure}[t]
  \centering
  \includegraphics{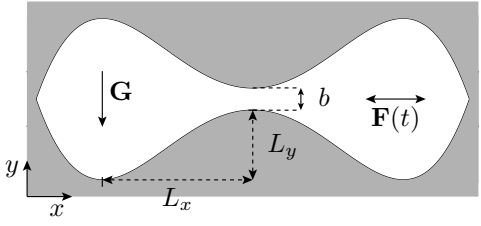}
  \caption{Schematic illustration of the two-dimensional
    structure confining the motion of the Brownian particles.
    The symmetric structure is defined by a quartic double well function,
    cf. Eq.~\eqref{eq:widthfunctions}, involving the geometrical
    parameters $L_{x}$, $L_{y}$ and $b$.
    Brownian particles are driven by a sinusoidal
    force $\vec{F}(t)$ along the longitudinal direction and a constant
    force $\vec{G}$ in the orthogonal direction.
  }
  \label{fig:well}
\end{figure}

The dynamics of a particle in a constrained geometry subjected to a
sinusoidal oscillating force $F(t)$ along the axis of the structure
and to a constant force $G$ acting along the orthogonal, or
transverse, direction can be described by means of the Langevin
equation written, in the overdamped limit, as
\begin{align}
  \label{eq:langevin}
  \gamma \, \frac{\mathrm{d}\vec{r}}{\mathrm{d} t} = - G\vec{e_y}
  - F(t)\vec{e_x} +\sqrt{\gamma \, k_{\mathrm{B}}T}\, \vec{\xi}(t)\, ,
\end{align}
where $\vec{r}$ denotes the position of the particle, $\gamma$ is
the friction coefficient, $\vec{e_x}$ and $\vec{e_y}$ the unit
vectors along $x$ and $y$-directions, respectively, and
$\vec{\xi}(t)$ is a Gaussian white noise with zero mean which obeys
the fluctuation-dissipation relation $\langle
\xi_{i}(t)\,\xi_{j}(t') \rangle = 2\, \delta_{ij}\, \delta(t - t')$
for $i,j = x,y$. The explicit form of the longitudinal force is
given by $F(t) = F_0 \sin( \Omega t )$ where $F_0$ is the amplitude
and $\Omega$ is the angular frequency of the sinusoidal driving.

In the presence of confinement, this equation has to be solved by
imposing reflecting (no-flow) boundary conditions
at the walls of the structure. For the 2D structure depicted in
Fig.~\ref{fig:well}, the walls are defined by
\begin{align}
  \label{eq:widthfunctions}
  w_{\mathrm{l}}(x) &= L_y \left(\frac{x}{L_x}\right)^4 -
  2\,L_y\left( \frac{x}{L_x}\right)^2 - \frac{b}{2} \, = -  w_{\mathrm{u}}(x)\, ,
\end{align}
where $ w_{\mathrm{l}}$ and $ w_{\mathrm{u}}$ correspond to the
lower and upper boundary functions, respectively. The characteristic
length $L_{x}$ refers to the distance between the bottleneck and the
position of maximal width, $L_{y}$ corresponds to the narrowing of
the boundary functions and $b$ to the remaining width at the
bottleneck, cf. Fig.~\ref{fig:well}. Consequently, the local width
of the structure reads: $2\,w(x) = w_{\mathrm {u}}(x) -
w_{\mathrm{l}}(x)$. This particular choice of the geometry is
intended to resemble the {\itshape classical setup} for SR in the
context of energetic barriers. In fact, in the limit of a
sufficiently large transverse force $G$, the particle is in practice
restricted to explore the region very close to the lower boundary of
the structure, recovering the effect of an energetic bistable
potential. For the sake of a dimensionless description, we measure
all lengths in units of $L_{x}$, i.e. ${\tilde{x}} = x/L_x$,
$\tilde{y} = y/L_{x}$ implying  ${\tilde {b}} = b/L_x$ and ${\tilde
{w}}_{\mathrm{l}}= w_{\mathrm{l}}/L_x = - {\tilde
  {w}}_{\mathrm{u}}$,
temperature in units of an arbitrary, but
irrelevant reference temperature $T_\mathrm{R}$
and time in units of $\tau = \gamma L_{x}^{2}/k_{\mathrm{B}}
T_\mathrm{R}$, that is, twice the time the particle takes to diffuse a
distance $L_{x}$ at temperature $T_\mathrm{R}$, i.e. ${\tilde {t}} = t/\tau$ and
$\tilde{\Omega} = \Omega \tau$. We scale forces by
$F_{\mathrm{R}}=\gamma L_x/\tau $, i.e. the orthogonal force reads ${\tilde {G}} = G/F_{\mathrm{R}}$
and the sinusoidal force ${\tilde {F}(\tilde {t})} =
F(t)/F_{\mathrm{R}}$. For better legibility, we shall omit
the tilde symbols in the following.
In dimensionless form the Langevin-equation~\eqref{eq:langevin} and
the boundary functions \eqref{eq:widthfunctions} read:
\begin{align}
  \label{eq:dllangevin}
  \frac{\mathrm{d}\vec{r}}{\mathrm{d} t} & = - G\vec{e_y}
  - F(t)\vec{e_x} +\sqrt{D}\, \vec{\xi}(t)\, ,\\
  \label{eq:dlboundaryfunctions}
  w_{\mathrm{l}}(x) & = -w_{\mathrm{u}}(x) =  \epsilon x^4 - 2\epsilon x^2
  - b/2 \, ,
\end{align}
where we defined the aspect ratio $\epsilon = L_y/L_x$ and the
rescaled temperature $D=T/T_{\mathrm{R}}$.

\section{Reduction of dimensionality}
\label{sec:rod}

Since the above mentioned dynamics given by Eq.~\eqref{eq:dllangevin} with the
boundary conditions could not be solved analytically,
we simplified the problem by
assuming equilibration in $y$- direction and thereby reducing the
dimensionality of the problem \cite{Jacobs,Zwanzig,Reguera_PRE,Percus}.

First, we consider the case in the absence of the periodic forcing,
i.e. $F(t)=0$. Then, the 2D dynamics is described by the following 2D Smoluchowski
equation  \cite{Risken,hanggithomas}
\begin{multline}
\label{eq:2dsmoluchuowski}
\frac{\partial}{\partial t} P(x,y,t) =
D\frac{\partial}{\partial x} e^{-U(x,y)/D}
\frac{\partial}{\partial x} e^{U(x,y)/D} P(x,y,t)\, \\ +
D\frac{\partial}{\partial y} e^{-U(x,y)/D}
\frac{\partial}{\partial y} e^{U(x,y)/D} P(x,y,t)\, ,
\end{multline}
with reflecting boundary conditions at the confining walls and where the
potential function is given by $U(x,y) = G\, y$. Since we are mainly
interested in the dynamics in $x$-direction, we introduce the marginal
probability density $P(x,t)$ which is obtained by integration over the transverse
coordinate:
\begin{align}
\label{eq:equlibrium}
P(x,t) = \int \mathrm{d}y \, P(x,y,t) \, .
\end{align}
On integrating Eq.~\eqref{eq:2dsmoluchuowski} over the transverse
direction, we get
\begin{multline}
\label{eq:1dsmoluchuowski}
\frac{\partial}{\partial t} P(x,t) = \\
D\frac{\partial}{\partial x} \int \, \mathrm{d}y \, \left\{e^{-U(x,y)/D}
  \frac{\partial}{\partial x} e^{U(x,y)/D} P(x,y,t) \right\} \, .
\end{multline}
Assuming local equilibrium in the $y$-direction, we define the
$x$-dependent effective
energy function $A(x)$ (omitting irrelevant constants) reading
\begin{align}
\label{eq:local}
e^{-A(x)/D} =
\int \mathrm{d}y \,e^{-U(x,y)/D} \, .
\end{align}
Consequently, the conditional local equilibrium probability
distribution of $y$ at a given $x$ becomes
\begin{align}
\label{eq:conditional}
\rho (y;x) = e^{-U(x,y)/D} \, e^{A(x)/D}\, ,
\end{align}
and is normalized for every $x$. As a result, the two dimensional probability
distribution can be approximately expressed as
\begin{align}
\label{eq:equilib}
P(x,y,t) \cong P(x,t) \rho (y;x) \, ,
\end{align}
and the kinetic equation for the marginal probability distribution,
cf. Eq.~\eqref{eq:1dsmoluchuowski}, becomes
\begin{align}
\label{eq:kineticeq}
\frac{\partial}{\partial t} P(x,t) \cong
D\frac{\partial}{\partial x} e^{-A(x)/D}
\frac{\partial}{\partial x} e^{A(x)/D} P(x,t) \, .
\end{align}
In the present case with a constant force $G$ in the negative
$y$-direction the potential function $A(x)$ reads, cf. Eq.~\eqref{eq:local}:
\begin{align}
\label{eq:effpotentialgen}
A(x) &= - D \ln \left[
\int_{w_\mathrm{l}(x)}^{w_\mathrm{u}(x)} \,
e^{-G\,y / D} \,\mathrm{d}y
\right] \nonumber \\
&= -D \ln \left[ \frac{D}{G} \left( e^{-G w_{\mathrm{l}}(x)/D} - e^{-G
      w_{\mathrm{u}}(x)/D} \right)\right]\, .
\end{align}
Making use of the symmetry of our considered structure,
i.e. $w_\mathrm{u}(x) = - w_\mathrm{l}(x)$ and the definition of the
half width function $w(x)=(w_{\mathrm{u}}-w_{\mathrm{l}})/2$, the potential function
turns into
\begin{align}
\label{eq:effpotential}
A(x)= -D\, \ln \left[\frac{2 D}{G}\,
\sinh\left(\frac{G w(x)}{D} \right)
\, \right]\, .
\end{align}
Eq.~\eqref{eq:kineticeq} can then be rewritten as
\begin{align}
\label{eq:fj}
\frac{\partial P(x,t)}{\partial t} =
     \frac{\partial}{\partial x}\left\{D \frac{\partial P}{\partial x} \, +
     A^{\prime}(x)\, P \right\} \, ,
\end{align}
with the potential function $A(x)$ given by
Eq.~\eqref{eq:effpotentialgen} (or by Eq.~\eqref{eq:effpotential} for $w_{\mathrm{l}}(x)=-w_{\mathrm{u}}(x)$)
and with the prime referring to the derivative with respect to $x$. In
general, after the coarse-graining the diffusion coefficient will
depend on the coordinate $x$, but since in our case $|w^{\prime}(x)| \ll 1$,
the correction can be safely neglected
\cite{Burada_PRE,Zwanzig,Reguera_PRE,Percus,Berezhkovskii2007}.

\begin{figure}[t]
  \centering
  \includegraphics{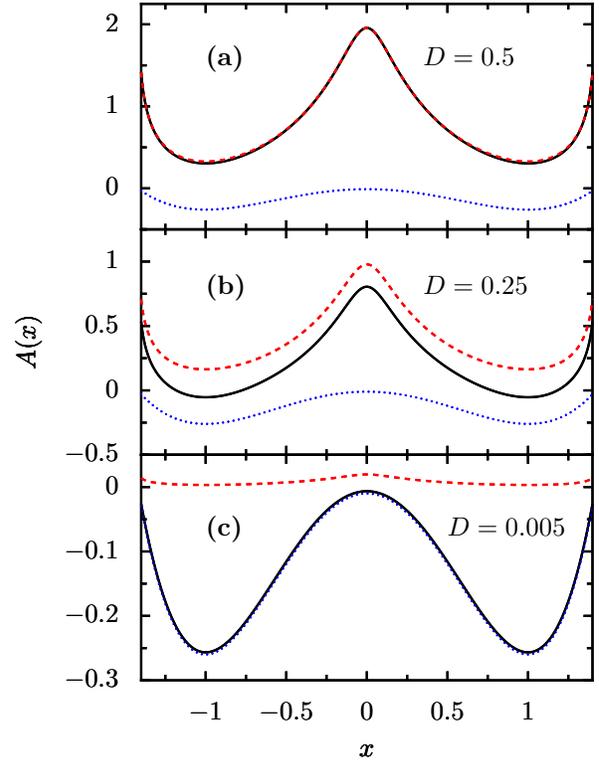}
  \caption{(Color online) The effective 1D potential $A(x)$ is
    depicted (solid black line) for $G=1$
    and for different noise strengths.
    For comparison the potential functions for the pure
    energetic case (dotted blue line), i.e. $A(x) = G w_{\mathrm{l}}(x)$,
    and for the pure entropic case (dashed red line),
    i.e. $A(x) = - D\ln[2 w(x)]$, are plotted as well.}
  \label{fig:potential}
\end{figure}

It is important to highlight that the potential $A(x)$ was not present
in the 2D Langevin dynamics, but arises
due to the entropic restrictions associated to the confinement. Then,
equation~\eqref{eq:fj} describes the motion of a Brownian particle in
a free energy potential of entropic nature, as $A(x)$
does not only depend on the energetic contribution of the force $G$, but also on the
temperature $D$ and the geometry of the structure in a non-trivial
way. For a structure like that depicted in Fig.~\ref{fig:well}, the
free energy $A(x)$ forms a double-well potential,
cf. Fig.~\ref{fig:potential}. As the width at the bottleneck of the
channel approaches zero, the potential $A(x)$ diverges.

It is worth to analyze the two limiting situations that can be
obtained depending on 
the transverse force $G$.
\begin{description}
\item[{\bf Energy-dominated situation:}]
For 
$G \gg 1$,
Eq.~\eqref{eq:effpotential} yields
$A(x) = G w_{\mathrm{l}}(x)$ (neglecting irrelevant constants) and the 1D kinetic
equation \eqref{eq:fj} becomes the standard
Fokker-Planck equation for Brownian motion in a purely energetic potential
whose shape resembles the lower boundary of the structure,
\begin{align}
  \label{eq:fk}
  \frac{\partial P(x,t)}{\partial t} =
  \frac{\partial}{\partial x}\left\{ D \frac{\partial P(x,t)}{\partial x} +
    G w_{\mathrm{l}}^{\prime}(x) P(x,t)\right\} \, .
\end{align}
\item[{\bf Entropy-dominated situation:}]
In the opposite limit,\newline 
i.e. for 
$G \ll 1$,
the effective potential is dominated by the purely entropic contribution
$A(x) = -D\,\ln[2\, w(x)]$ and the kinetic equation
turns into the Fick-Jacobs equation \cite{Jacobs},
\begin{align}
  \label{eq:ent}
  \frac{\partial P(x,t)}{\partial t} =
  \frac{\partial}{\partial x}\bigg\{ D \frac{\partial P(x,t)}{\partial
    x} - D \frac{w'(x)}{w(x)}P(x,t) \bigg\} \, .
\end{align}
\end{description}

\section{Two-State approximation}
\label{sec:ts}

It is instructive to analyze the occurrence of stochastic resonance in
the context of the two-state approximation \cite{McNamara}.
Accordingly, the 1D kinetics given by
Eq.~\eqref{eq:fj} can be approximately mapped into a two-state
system with the two states corresponding to the two wells of the
symmetric effective potential $A(x)$. An estimate for the
transition rates
could be obtained by applying the Mean-First-Passage-Time (MFPT)
approach \cite{hanggi}.

\begin{figure}[t]
  \centering
  \includegraphics{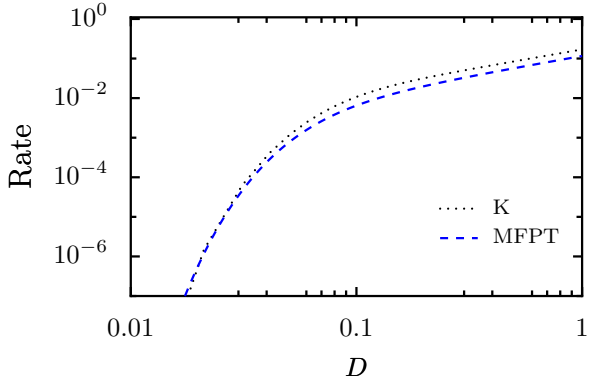}
  \caption{(Color online) The transition rate for the escape from one basin to the other
    obtained after applying a Two-State approximation of the 1D problem,
    cf. Eq.~\eqref{eq:rmfpt} and Eq.~\eqref{eq:kr-both}, is
    depicted as a function of the thermal noise $D$ for $G=1.0$.
    The 2D structure is defined by the boundary function given
    in Eq.~\eqref{eq:dlboundaryfunctions} with $\epsilon=1/4$ and $b=0.02$.
    Here, the Kramers-Smoluchowski rate ($K$), i.e. $r_{\mathrm{K}}$,
    converges to the rate given by the inverse mean first passage time (MFPT) in the
    low noise limit.}
  \label{fig:rate}
\end{figure}

\subsection{Mean First Passage Time approach}

In order to calculate the transition rate from one state to the
other, one evaluates the inverse of the MFPT to reach a potential
minimum after starting out from the other minimum of the symmetric
and bistable potential $A(x)$. Then, the transition rate is given by
\begin{align}
  \label{eq:rmfpt}
  r_{\mathrm{MFPT}}( D, G) = \frac{1}{T_{1}( -1 \to 1)}\, ,
\end{align}
where $T_{1}(-1 \to 1)$ is the first moment of the
first passage time distribution for reaching $x=1$ starting out at $x=-1$.
The $n$th moment of the first passage time distribution obeys
the following recurrence relation \cite{hanggi}

\begin{align}
  \label{eq:st-average}
  T_n (-1 \to 1) & := \langle t^n(-1 \to 1)\rangle
  \nonumber \\
  & = \frac{n}{D} \int_{-1}^{1} \mathrm{d}x \, e^{A(x)/D}
  \int_{-\infty}^{x} \mathrm{d}y \,
  e^{- A(y)/D}
  \nonumber \\
  & \times \langle t^{n-1}( y \to 1 )\rangle
\end{align}
for $n \in \mathbb{N}$ and with $T_0(a \rightarrow b) = 1$
for arbitrary a and b.
Accordingly, within the one dimensional approximation,
cf. Eq.~\eqref{eq:fj}, the mean
first passage time (i.e., $n = 1$)
for the potential function $A(x)$,
cf Eq.~\eqref{eq:effpotential}, reads
\begin{align}
\label{eq:1st}
T_1 & = \frac{1}{D} \int_{-1}^{1} \mathrm{d}x\, e^{A(x)/D} \int_{-x_\mathrm{l}}^{x}
\mathrm{d}y \,e^{- A(y)/D} \nonumber \\
& = \frac{1}{D} \int_{-1}^{1} \mathrm{d}x\, \mathrm{csch}\left(\frac{G w(x)}{D}\right)
\int_{-x_\mathrm{l}}^{x} \mathrm{d}y\, \sinh\left(\frac{G w(y)}{D}\right) \, ,
\end{align}
where $x_\mathrm{l}$ is the left limiting value at which the
boundary function vanishes. Eq.~\eqref{eq:1st} can be evaluated
using a steepest descent approximation leading to the commonly known
Kramers-Smoluchowski rate.

\subsection{Kramers-Smoluchowski rate}

For a potential $A(x)$ with a barrier height $\Delta A \gg D$
the escape rate of an overdamped Brownian particle from one well
to the other in the presence of thermal
noise, and in the absence of a force, is given by the
Kramers-Smoluchowski rate \cite{McNamara,hanggi,kramers,Jung91},
reading in dimensionless units,
\begin{align}
\label{eq:krate}
r_\mathrm{K}(D) = \frac{\sqrt{A^{\prime \prime}(x_\mathrm{min})|A^{\prime \prime}
(x_\mathrm{max})|}}{2\pi}\,
\exp\left(\frac{-\Delta A}{D}\right) \, ,
\end{align}
where $A^{\prime \prime}$ is the second derivative of the
effective potential function with respect to $x$,
and $x_\mathrm{max}$ and $x_\mathrm{min}$ indicate the position
of the maximum and minimum of the
symmetric potential, respectively.

For the potential given by Eq.~\eqref{eq:effpotential} and the
shape defined by Eq.~\eqref{eq:dlboundaryfunctions},
the corresponding Kramers-Smoluchowski rate for transitions from one
basin to the other reads \cite{Burada_PRL}
\begin{align}
  \label{eq:kr-both}
  r_\mathrm{K}(D) = \frac{G\,\epsilon}{\pi} \,
\frac{\sqrt{2\,\sinh\left(\frac{G b}{D}\right)\,
\sinh\left(\frac{G( b + 2 \epsilon)}{D}\right)}}
{{\sinh}^{2}\left(\frac{G( b + 2 \epsilon)}{2 D}\right)}
\end{align}

Note that the Kramers-Smoluchowski approximation yields good results
for barrier heights $\Delta A$ much larger than the thermal energy
that in the present scaling is given by $D$. In fact, for $\Delta A
\gg D$ the Kramers-Smoluchowski rate $r_\mathrm{K}$ approximates
accurately the rate $r_{\mathrm{MFPT}}$ evaluated numerically
from the MFPT expression ~\eqref{eq:1st}, as depicted in
Fig.~\ref{fig:rate}.

\section{Role of the transverse force $\vec{G}$}
\label{sec:rotof}

In section \ref{sec:rod} we introduced the Fick-Jacobs equation to
approximatively describe the Brownian motion in a 2D structure like
that depicted in Fig.~\ref{fig:well} using a simplified 1D modeling
with an effective bistable potential. This potential exhibits a
barrier the particle has to overcome noisily in order to make a
transition from one well to the other. For a sinusoidal driving
force applied in $x$-direction, i.e. $F(t) = F_{0} \, \sin( \Omega
t)$, a synchronization effect between the oscillatory forcing and
the noise-induced transitions over the entropic potential barrier
takes place and was reported previously in Ref.~\cite{Burada_PRL}.
Under these circumstances and in the presence of a finite
orthogonal, i.e. transverse force $G$, increasing the noise
level results in a noise-amplified response signal. The improvement
of the response is quantified in the following by the spectral
amplification factor $\eta$, which is the ratio of the power stored
in the response of the system at the driving frequency $\Omega$ and
the power of the sinusoidal driving signal. The occurrence of the
{\itshape Entropic Stochastic Resonance} effect manifests in the
presence of an optimal dose of noise for which the spectral
amplification is maximal \cite{Burada_PRL}.

\subsection{Two-State modeling}

 It is straight forward to derive an
analytic expression for the spectral amplification within a
two-state modeling. The sinusoidal driving modulates the transition
rates, which are given either by $r_{\mathrm{MFPT}}$ or
$r_{\mathrm{K}}$ (cf. Sec.~\ref{sec:ts}). Within a first order
perturbation theory in the ratio of driving amplitude $F_{0}$ and
noise level $D$, it is possible to find a closed expression for the
response of the two-state system and accordingly for the spectral
amplification factor that reads \cite{gammaitoni,Jung91}:
\begin{align}
  \label{eq:amplification}
  \eta = \frac{1}{D^2}\, \frac{4\, r_\mathrm{Rate}^{2}(D)}{4\,r_\mathrm{Rate}^{2}(D) +
    \Omega^{2}} 
  \, ,
\end{align}
where $r_{\mathrm{Rate}}$ is the transition rate of the unperturbed
two-state system and is given by Eq.~\eqref{eq:rmfpt}, or approximately by
Eq.~\eqref{eq:krate}.

\begin{figure}[t]
  \centering
  \includegraphics{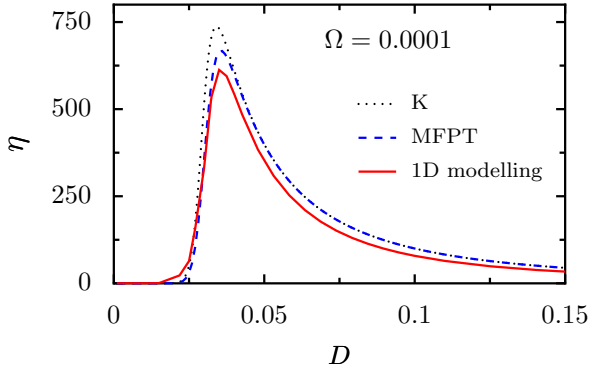}
  \caption{(Color online) A detailed comparison of the behavior of
   the spectral amplification factor obtained numerically from the 1D modeling (solid line),
    cf. Eq.\eqref{eq:fj-full} and Eq.\eqref{eq:samplification}, and
    within the two-state approximation (Eq.\eqref{eq:amplification}),
    using either the Kramers-Smoluchowski
    rate (K) (dotted line) or the rate obtained directly from the mean first passage time (MFPT) (dashed-line).
    Here, the transverse force is $G = 1.0$, and the structure is defined
    by Eq.~\eqref{eq:dlboundaryfunctions} with an aspect ratio $\epsilon = 1/4$, and
    a bottleneck width $b = 0.02$.}
  \label{fig:comparision}
\end{figure}

\subsection{1D modeling}

Avoiding the approximations involved in the two-state modelling, the
system's response could also be obtained directly from the numerical
integration of the 1D kinetic equation. In the presence of an
oscillating force $F(t)$ in $x-$direction there is an additional
contribution to the effective potential in Eq.~\eqref{eq:fj} and the
1D kinetic equation in dimensionless units  reads \cite{Burada_PRL}
\begin{align}
  \label{eq:fj-full}
  \frac{\partial P(x,t)}{\partial t} =
  \frac{\partial}{\partial x}\left\{D \frac{\partial P}{\partial x}  +
  \textbf{(} A^{\prime}(x) - F(t) \textbf{)}\,P \right\} \, .
\end{align}

By spatial discretization, using a Chebyshev collocation method, and
employing the method of lines, we reduced the kinetic equation to a
system of ordinary differential equations, which was then solved using a
backward differentiation formula method \cite{nag}. As a result, we obtained the time
dependent probability distribution $P(x,t)$ and, from that,
the time-dependent average position defined as
\begin{align}
  \label{eq:meanx}
  \langle x(t) \rangle = \int x \, P(x,t) \mathrm{d} x \, ,
\end{align}
which was computed in the long-time limit. After a Fourier-expansion
of $\langle x(t) \rangle $ one finds the amplitude $M_1$ of the
first harmonic of the system's response. Hence, the spectral
amplification $\eta$ \cite{Jung91} for the fundamental oscillation
is evaluated as
\begin{align}
\label{eq:samplification}
\eta = \left[\frac{M_1}{F_0} \right]^2\, .
\end{align}

The spectral amplification $\eta$ depicts a bell-shaped behavior,
cf. Fig.~\ref{fig:comparision}, indicating the existence of a
{\itshape Stochastic Resonance} effect: there is a maximum  of the
spectral amplification at an optimal value of noise. The qualitative
behavior is captured by the two-state approximation as well. The
prediction of the two-state modelling with rates evaluated directly
from MFPT is closer to the 1D modelling than those obtained by
making use of the Kramers-Smoluchowski approximated rate, as
illustrated in Fig.~\ref{fig:comparision}. This is due to the fact
that the condition $\Delta A \gg D$ is not always fulfilled in this
case. In fact for the entropy-dominated situation, $\Delta A$ and
$D$ are of the same order of magnitude.

Next we present the results of numerical simulations of the full
(2D) problem. By doing so we demonstrate that the ESR effect is
robust and  not just an artifact of the reduction procedure.

\begin{figure}[t]
  \centering
  \includegraphics{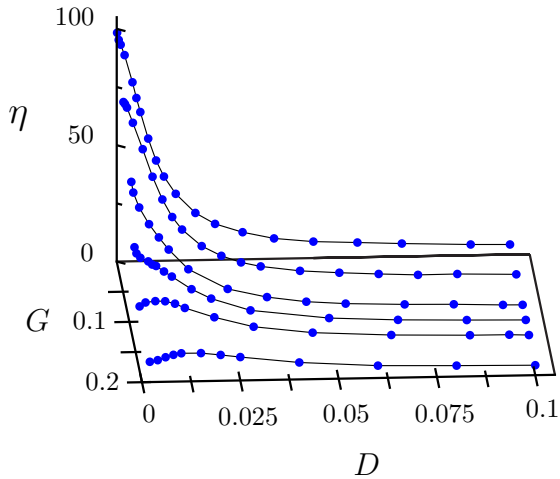}
  \caption{(Color online) The spectral amplification obtained from 2D
    simulations is depicted as a function of both the
    transverse force $G$ and the noise strength $D$ for an
    input signal frequency $\Omega = 0.1$ and amplitude $F_0 = 0.1$.
    The shape of the two-dimensional channel is defined by
    the dimensionless function
    $w(x) = -\epsilon x^4 + 2\,\epsilon x^2 + b/2$
    with the aspect ratio $\epsilon = 1/4$ and bottleneck width $b=0.02$.
    The symbols connected by lines correspond to the same
    $G$-values. In particular, these are $G=0.0,\, 0.05, \, 0.1, \,
    0.125, \, 0.15$ and $0.2$.
    In the deterministic limit, i.e $D \to 0$,
    the spectral amplification reaches
    a limiting value ($1/\Omega^2$) for $G = 0$. For $G=0$ the
    effect of {\itshape Entropic Stochastic Resonance} disappears.}
  \label{fig:3dgraph}
\end{figure}

\subsection{2D modelling}

The accuracy of the reduced one-dimensional
kinetic description can be examined by
comparing the results with those obtained by Brownian dynamic simulations,
performed by integration of the overdamped 2D Langevin
equation \eqref{eq:langevin}.
The simulations were carried out
using the standard stochastic Euler-algorithm.

\begin{figure}[t]
  \centering
  \includegraphics{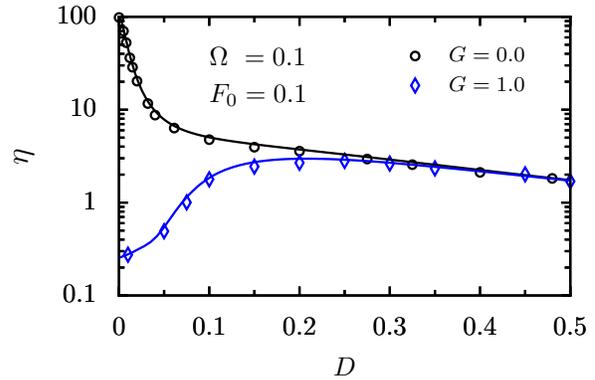}
  \caption{(color online)
    The dependence of the spectral amplification
    $\eta$ on noise level $D$,
    at two different values of the transverse force $G$,
    at a constant input frequency and amplitude.
    The symbols correspond to the results of the Langevin simulations
    for the two-dimensional channel whose shape is defined by
    $w(x) = -\epsilon x^4 + 2\,\epsilon x^2 + 0.01$
    with an aspect ratio $\epsilon = 1/4$,
    whereas the lines are the results of the numerical integration of
    the 1D kinetic equation \eqref{eq:fj-full}.
    In the deterministic limit, i.e $D \to 0$,
    the spectral amplification reaches
    the limiting value ($1/\Omega^2$) for $G = 0$.}
  \label{fig:gravity}
\end{figure}

The resulting amplification factor as a function of the value of the
transverse force and the noise strength are plotted in
Fig.~\ref{fig:3dgraph}. For a finite transverse force $G$ the
spectral amplification exhibits a peak at an optimum value of the
noise strength which is indicative of the effect of {\itshape
Entropic Stochastic Resonance}. However, for vanishing transverse
force $G$ the spectral amplification does not exhibit any peak
and decays monotonically with increasing noise level, i.e. the ESR -
effect is not observed. In the deterministic limit, i.e. $D \to 0$,
the spectral amplification reaches the limit value $\eta =
1/\Omega^2$ for $G=0$, as shown in Fig.~\ref{fig:3dgraph}. We remark
that the later is only true when the amplitude of the system's
response, which is the ratio of input signal amplitude to input
signal frequency, is smaller than the value $x_{l}$, which is the
limiting value at which the boundary function vanishes. It is also
worth to point out that starting out from a finite $G$-value the
position of the ESR peak shifts towards smaller noise strengths as
$G$ decreases while the maximum value increases, cf.
Fig.~\ref{fig:3dgraph}. A comparison between the 1D-modelling and
the full 2D simulation is depicted in Fig.~\ref{fig:gravity}. The
2D simulation results convincingly corroborate the validity of the
modelling within the 1D Fick-Jacob approximation, see also below.

Thus, we detect a non-monotonic behavior of the spectral
amplification only for finite values of the orthogonal force $G \neq
0$, while for $G=0$ the spectral amplification decays monotonically.
In other words, for the dynamics and situation considered in
Sec.~\ref{sec:cbm} the occurrence of the ESR-effect requires a
non-vanishing orthogonal force $G$ which contributes within the 1D
modelling to the effective entropic potential.

\section{Role of the shape of the structure}
\label{sec:rotsots}

Fig.~\ref{fig:gravity} depicts the main findings for the occurrence
of the ESR-effect in an effective bistable configuration with two
large basins connected by a small bottleneck, cf.
Fig.~\ref{fig:well}. Namely: (i) the occurrence of the ESR-effect
due to the interplay of a finite value of the orthogonal force $G$,
and (ii) the ability of the 1D Fick-Jacobs approximation to
reproduce very accurately the results of the full 2D problem
\cite{Burada_PRE,Burada_PRL}.

\begin{figure}[t]
  \centering
  \includegraphics{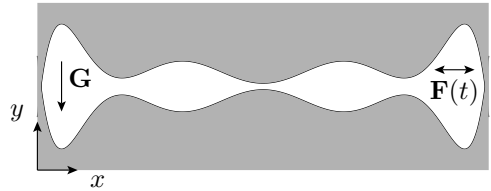}
  \caption{Schematic illustration of the two-dimensional
    structure confining the motion of the Brownian particles.
    The shape defined by the dimensionless function
    $w(x) = 12.5 x^8 - 27.5 x^6 + 18.21 x^4 - 3.92 x^2 - 0.01$.}
  \label{fig:well-2}
\end{figure}

Next, in order to investigate the impact of the shape of the
structure on the ESR behavior we have considered yet another channel
geometry: It is similar to the one depicted in Fig.~\ref{fig:well}
but exhibits two intermediate wider bottlenecks which are connected
via a much narrower bottleneck, see in Fig.~\ref{fig:well-2}. The
geometric shape of the structure is defined by the dimensionless
width function
\begin{align}
  \label{eq:shape-2}
  w(x) = 12.5 x^8 - 27.5 x^6 + 18.21 x^4 - 3.92 x^2 - 0.01 \, .
\end{align}

Following the same analysis performed in Sec.~\ref{sec:rotof}, we
obtained the
spectral amplification within the 1D modelling and compared it with
results obtained from 2D numerical simulation of the Langevin equation
\eqref{eq:dllangevin} in the new 2D structure defined by
Eq.~\eqref{eq:shape-2}. Again, the effective potential function $A(x)$ obtained
from the Fick-Jacobs approximation exhibits a large potential barrier
separating the two basins each of which is additionally separated by a
smaller potential barrier into two wells. In fact, the construction of
the alternative geometric structure was done in such a way to reveal
the existence of a main entropic barrier controlling the transitions
between the two main basins.

\begin{figure}[t]
  \centering
  \includegraphics{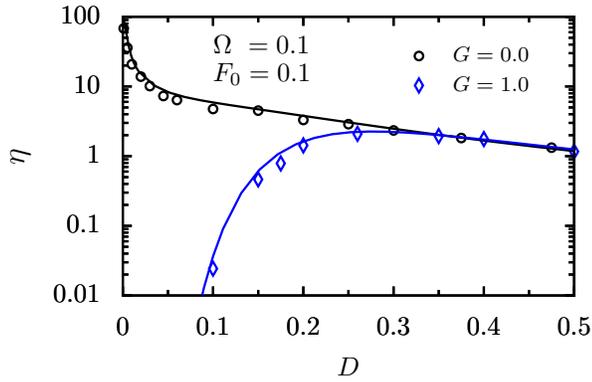}
  \caption{(Color online)
    The dependence of the spectral amplification
    $\eta$ on noise level $D$, at two different values of
    the transverse force $G$, at a constant input
    frequency and amplitude
    for the channel illustrated in Fig.~\ref{fig:well-2}).
    The symbols correspond to the results of the Langevin simulations
    for the two-dimensional structure whose shape is defined by
    $w(x) = 12.5 x^8 - 27.5 x^6 + 18.21 x^4 - 3.92 x^2 - 0.01$,
    whereas the lines are the results of the numerical integration of
    the 1D kinetic equation \eqref{eq:fj-full}.
    Like in Fig.~\ref{fig:gravity}, in the deterministic limit ($D \to 0$),
    the spectral amplification reaches
    the limiting value ($1/\Omega^2$) for $G = 0$.}
  \label{fig:SR-2}
\end{figure}

The behavior of spectral amplification as a function of the noise
strength for the new channel defined by Eq.~\ref{eq:shape-2}, and
for two different values of the transverse force is depicted in
Fig.~\ref{fig:SR-2}. Interestingly, we can observe the existence of
Entropic Stochastic Resonance even in the new structure, when the
transverse force is present in the system. Comparing the results of
Fig.~\ref{fig:SR-2} for the channel depicted in
Fig.~\ref{fig:well-2} with those in Fig.~\ref{fig:gravity}, one can
observe that the ESR peak appears at higher values of the noise
strength. In addition, the enhancement of the signal upon increasing
the noise is more pronounced in the new structure.  These results
suggest that by a proper design of the geometry of the channel it
would be possible to significantly enhance and optimize the response
of a confined system. This would be specially important in
biological systems, where the noise (i.e. the temperature) is a
variable that can neither be arbitrarily chosen nor eliminated.
Finally, it is also worth stressing that the Fick-Jacobs
approximation still holds nicely in this case since the 2D numerical
simulation results are in very good agreement with those obtained by
1D modelling.

Overall, the results indicate that the existence of a small bottleneck
separating two basins and
forming an effective entropic potential barrier, leads robustly to the
occurrence of an {\itshape Entropic Stochastic Resonance} effect.


\section{Conclusions}

We have shown that unevenness may be the origin of many resonant
phenomena in small-scale systems. The constrained motion of a
Brownian particle in a region limited by irregular boundaries
impedes the access of the particle to certain regions of space
giving rise to entropic effects that can effectively control the
dynamics. The interplay between the noise present in the system, the
external modulation and the entropic effects results in an entropic
stochastic resonance. However, the presence of a transverse force
$G$ is crucial to observe this resonant behavior.
This ESR is genuine of small scale systems where confinement yields
entropic effects. The occurrence of ESR depends on the shape of the
channel and can then be controlled by it. Thus understanding the
role of noise and confinement in these systems does provide the
possibility for a design of stylized channels wherein  response and
transport become efficiently optimized.

\bigskip
This work has been supported by the DFG via research
center, SFB-486, project A10, the
Volkswagen Foundation (project I/80424), the German Excellence
Initiative via the \textit {Nanosystems Initiative Munich} (NIM), and
the DGCyT of the Spanish government through grant No. FIS2005-01299.


\begin{thebibliography}{}

\bibitem{gammaitoni} L. Gammaitoni, P. H\"anggi, P. Jung, F. Marchesoni,
  Rev. Mod. Phys. \textbf{70}, 223 (1998)

\bibitem{bulsaraJS}
A. Bulsara, P. H\"anggi, F. Marchesoni, F. Moss, M. Shlesinger,
J. Stat. Phys.  {\bf 70}, 1 (1993)

\bibitem{PT_SR}
  A.R. Bulsara, L. Gammaitoni, Phys. Today {\bf 49} (3), 39 (1996)

\bibitem{chemphyschem} P. H\"anggi,
ChemPhysChem {\bf 3}, 285 (2002)

\bibitem{vilar_mono}J.M.G. Vilar, J.M. Rub\'i,
  Phys. Rev. Lett. {\bf 77}, 2863 (1996);
  J.M.G. Vilar, J.M. Rub\'i,
  Phys. Rev. Lett. {\bf 78}, 2886 (1997)

\bibitem{Lutz99}
  V.S. Anishchenko, A.B. Neiman, F. Moss, L. Schimansky-Geier,
  Phys. Usp. {\bf 42}, 7 (1999)

\bibitem{Schmid01} G. Schmid, I. Goychuk, P. H\"anggi,
  Europhys. Lett. {\bf 56}, 22 (2001)

\bibitem{BuchSR}
  T. Wellens, V. Shatokhin, A. Buchleitner, Rep. Prog. Phys. {\bf 67}, 45 (2004)

\bibitem{Yasuda08} H. Yasuda, T. Miyaoka, J. Horiguchi, A. Yasuda,
  P. H{\"a}nggi, Y. Yamamoto, 
  Phys. Rev.  Lett. {\bf 100}, 118103 (2008)

\bibitem{scholarpediaSR}
  C.R. Nicolis, G. Nicolis, Scholarpedia, {\bf 2}(11), 1474 (2007)

\bibitem{GoychukSRa}
  I. Goychuk, P. H\"anggi, Phys. Rev. Lett. {\bf 91}, 070601 (2003)

\bibitem{GoychukSRb}
I. Goychuk, P. H{\"a}nggi, J. L. Vega, S. Miret Artes, Phys. Rev. E {\bf 71}, 061906 (2005)

\bibitem{hille}B. Hille, {\it Ion Channels of Excitable Membranes}
  (Sinauer, Sunderland, 2001)

\bibitem{zeolites}R.M. Barrer, {\it Zeolites an Clay Minerals as Sorbents
    and Molecular Sieves} (Academic Press, London, 1978)

\bibitem{liu}L. Liu, P. Li, S.A. Asher,
  Nature {\bf 397}, 141 (1999)

\bibitem{berzhkovski}A.M. Berezhkovskii, S.M. Bezrukov,
  Biophys. J. {\bf 88}, L17(2005)


\bibitem{Reguera_PRL}D. Reguera,
  G. Schmid, P.S. Burada, J.M. Rub\'i, P. Reimann, P. H\"anggi,
  Phys. Rev. Lett. {\bf 96}, 130603 (2006)

\bibitem{Burada_PRL} P.S. Burada, G. Schmid,
  D. Reguera, M.H. Vainstein, J.M. Rubi, P. H\"anggi,
  Phys. Rev. Lett. \textbf{101}, 130602 (2008)

\bibitem{hanggithomas} P. H\"anggi, H. Thomas, Phys. Rep. \textbf{88}, 207 (1982)

\bibitem{Risken} H. Risken, {\it The Fokker-Planck equation}, 2nd ed. (Springer, Berlin, 1989)


\bibitem{Burada_PRE}P.S. Burada,
  G. Schmid, D. Reguera, J.M. Rub\'i, P. H\"anggi,
  Phys. Rev. E {\bf 75}, 051111 (2007);
  P.S. Burada, G. Schmid, P. Talkner, P. H\"anggi, D. Reguera, J.M. Rub\'i,
  BioSystems {\bf 93}, 16 (2008)

\bibitem{Jacobs}M. H. Jacobs, \emph{Diffusion Processes} (Springer, New York, 1967)

\bibitem{Zwanzig}R. Zwanzig, J. Phys. Chem. {\bf 96}, 3926 (1992)

\bibitem{Reguera_PRE}D. Reguera, J.M. Rub\'i,
  Phys. Rev. E {\bf 64}, 061106 (2001)

\bibitem{Percus} P. Kalinay, J.K. Percus,
  Phys. Rev. E {\bf 74}, 041203 (2006)

\bibitem{Berezhkovskii2007} A. M. Berezhkovskii,
  M.A.  Pustovoit, S.M. Bezrukov,
  J. Chem. Phys. {\bf 126}, 134706 (2007)

\bibitem{McNamara} B. McNamara, K. Wiesenfeld,
  Phys. Rev. A {\bf 39}, 4854 (1989)

\bibitem{hanggi} P. H\"anggi, P. Talkner, M. Borkovec,
  Rev. Mod. Phys. {\bf 62}, 251 (1990)

\bibitem{kramers} H. Kramers,
  Physica (Utrecht) {\bf 7}, 284 (1940)


\bibitem{Jung91}
  P. Jung, P. H\"anggi, Phys. Rev. A {\bf 44}, 8032 (1991);
  {\it ibid}, Europhys. Lett. {\bf 8}, 505 (1989)

\bibitem{nag} \emph{NAG Fortran Library Manual, Mark 20}
   (The Numerical Algorithm Group Limited, Oxford, England, 2001)


\end{thebibliography}
\end{document}